\documentclass[twocolumn,english,prx,showpacs,preprintnumbers,altaffilletter]{revtex4}
\usepackage[T1]{fontenc}
\usepackage[latin9]{inputenc}
\usepackage{color}
\usepackage{textcomp}
\usepackage{amsmath}
\usepackage{amssymb}
\usepackage{graphicx}

\makeatletter
\@ifundefined{textcolor}{}
{%
 \definecolor{BLACK}{gray}{0}
 \definecolor{WHITE}{gray}{1}
 \definecolor{RED}{rgb}{1,0,0}
 \definecolor{GREEN}{rgb}{0,1,0}
 \definecolor{BLUE}{rgb}{0,0,1}
 \definecolor{CYAN}{cmyk}{1,0,0,0}
 \definecolor{MAGENTA}{cmyk}{0,1,0,0}
 \definecolor{YELLOW}{cmyk}{0,0,1,0}
}

\usepackage{dcolumn}\usepackage{bm}

\usepackage{babel}

\usepackage{babel}

\makeatother

\usepackage{babel}
\begin{document}

\title{Supercurrent Spectroscopy of Andreev States}

\author{L.\ Bretheau$^{*}$, \c{C}. \"{O}. Girit$^{*}$, C.\ Urbina, D.\ Esteve
and H.\ Pothier}

\email[Corresponding author~: ]{hugues.pothier@cea.fr}

\thanks{$^{*}$L.\ Bretheau and \c{C}. \"{O}. Girit contributed equally to this
work.}

\affiliation{Quantronics Group, Service de Physique de l'\'{E}tat Condens\'{e} (CNRS,
URA\ 2464), IRAMIS, CEA-Saclay, 91191 Gif-sur-Yvette, France}

\date{\today}
\begin{abstract}
We measure the excitation spectrum of a superconducting atomic contact.
In addition to the usual continuum above the superconducting gap,
the single particle excitation spectrum contains discrete, spin-degenerate
Andreev levels inside the gap. Quasiparticle excitations are induced
by a broadband on-chip microwave source and detected by measuring
changes in the supercurrent flowing through the atomic contact. Since
microwave photons excite quasiparticles in pairs, two types of transitions
are observed: Andreev transitions, which consists of putting two quasiparticles
in an Andreev level, and transitions to odd states with a single quasiparticle
in an Andreev level and the other one in the continuum. In contrast
to absorption spectroscopy, supercurrent spectroscopy allows detection
of long-lived odd states. 
\end{abstract}


\maketitle

\section{INTRODUCTION}

\begin{figure}
\includegraphics[clip,width=1\columnwidth]{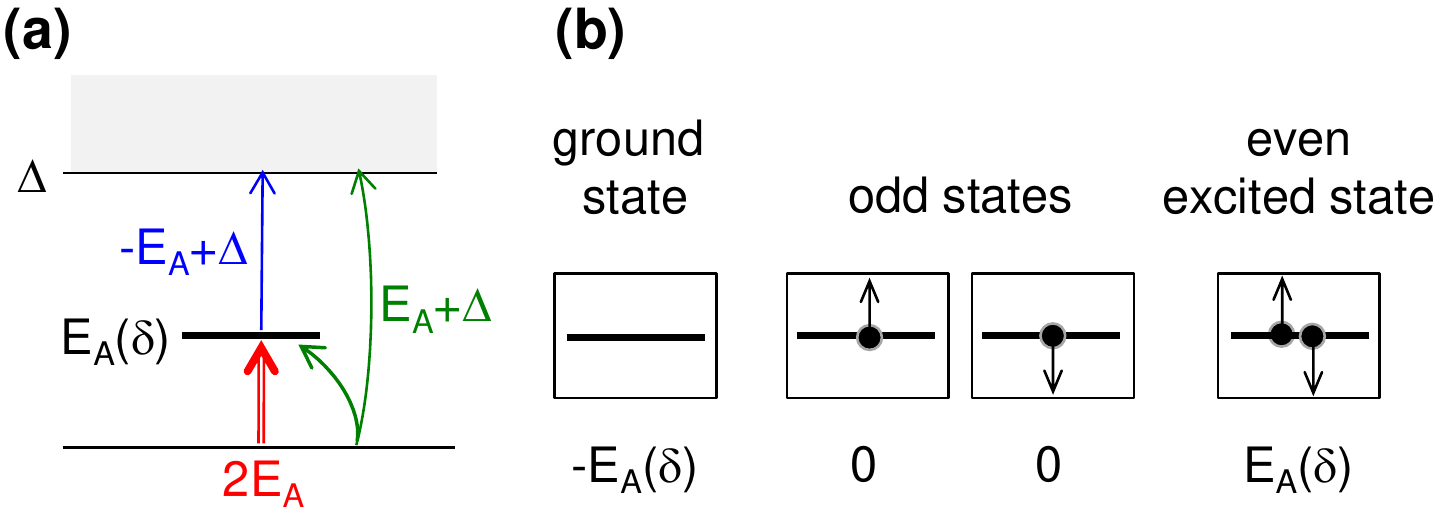}\caption{\textbf{(}\textbf{\textcolor{black}{a)}} The single particle excitation
spectrum for each channel of a weak link consists of a doubly-degenerate
Andreev level at energy $E_{A}\left(\delta\right),$ and a continuum
of states at energies larger than the superconducting gap $\Delta.$
Arrows indicate transitions that can be induced by microwaves. The
four possible occupations of the Andreev level are shown in\textbf{
(}\textbf{\textcolor{black}{b)}}: they correspond to the ground state,
the two odd states, and the excited pair state, with energies $-E_{A},$
$0$ and $E_{A}.$}

\label{principle} 
\end{figure}

The Josephson supercurrent between two superconductors \cite{Josephson}
exists in all types of weak links, including tunnel junctions, constrictions,
molecules, and normal metals. Weak links differ one from another by
their quasiparticle excitation spectrum, which is determined by the
length of the weak link and the transmission probabilities $\tau_{i}$
for electrons through each conduction channel $i$. In tunnel junctions,
$\tau_{i}\ll1$, and all excitations conserving electron parity require
energies at least equal to $2\Delta,$ where $\Delta$ is the superconducting
gap energy. With energy $2\Delta,$ a pair can be broken and two quasiparticles
created at the gap energy $\Delta.$ This is the same situation as
in a bulk superconductor. In contrast, the excitation spectrum of
weak links that have well transmitted channels, such as superconducting
constrictions, contains sub-gap spin-degenerate Andreev levels (Andreev
doublets) (see Fig.~1a). The energy of the Andreev level associated
with one channel with transmission $\tau$ in a short weak link is
$E_{A}=\Delta\sqrt{1-\tau\sin^{2}\left(\delta/2\right)}$ \cite{Beenakker,Bagwell},
with $\delta$ the superconducting phase difference across the weak
link. The lowest energy excitation that conserves electron parity,
the ``Andreev transition'', has an energy $2E_{A}:$ it consists
in the creation of two quasiparticles in the Andreev level (red double
arrow in Fig.~\ref{principle}a), which can be thought of as the
excitation of a pair localized at the weak link \cite{Nature,LBThesis}.
We recently reported microwave spectroscopy of this Andreev transition
in superconducting atomic contacts \cite{Nature}. There is a second
type of excitation, with energy at least $E_{A}+\Delta,$ in which
a localized Andreev pair is broken into one quasiparticle in the Andreev
level and one in the continuum (green arrows in Fig.~\ref{principle}a).
This process was addressed theoretically in recent works \cite{Bergeret,Bergeret2,Kos},
but has never been observed experimentally. Here, we describe how
``supercurrent spectroscopy'' reveals all possible transitions involving
Andreev states (Fig.~\ref{principle}). This method is based on measuring
the supercurrent through a weak link and detecting changes induced
by microwave excitation \cite{resonant}.

The supercurrent in a weak link results from the phase-dependence
of its energy, which depends on the occupation of the Andreev levels.
In the case of a single channel (see Fig.~1b), the ground state,
with energy $-E{}_{A}$, carries a supercurrent $-I_{A}=-\varphi_{0}^{-1}\left(\partial E_{A}/\partial\delta\right)$.
The energy of an odd state, with a single Andreev excitation, is zero:
it carries no supercurrent. The excited pair state with two Andreev
excitations has energy $+E_{A},$ and therefore carries a supercurrent
$+I_{A}.$ This difference in the supercurrent associated with 0,
1 or 2 Andreev excitations is the elementary phenomenon to understand
microwave- or voltage-induced variations of the supercurrent \cite{Fuechsle,baselmans}
as well as the current response to an a.c. field \cite{Dassonneville}
in diffusive superconductor\textendash{}normal-metal\textendash{}superconductor
junctions \cite{SNS}. Measurements of the phase dependence of the
supercurrent through atomic contacts with a few conduction channels
revealed the spontaneous excitation to odd states and allowed characterization
of their dynamics \cite{Zgirski}. We use here the dependence of the
supercurrent on the occupation of the Andreev doublet to reveal the
complete excitation spectrum of an elementary, generic weak link:
an atomic contact.

\begin{figure*}
\includegraphics[clip,width=0.8\textwidth]{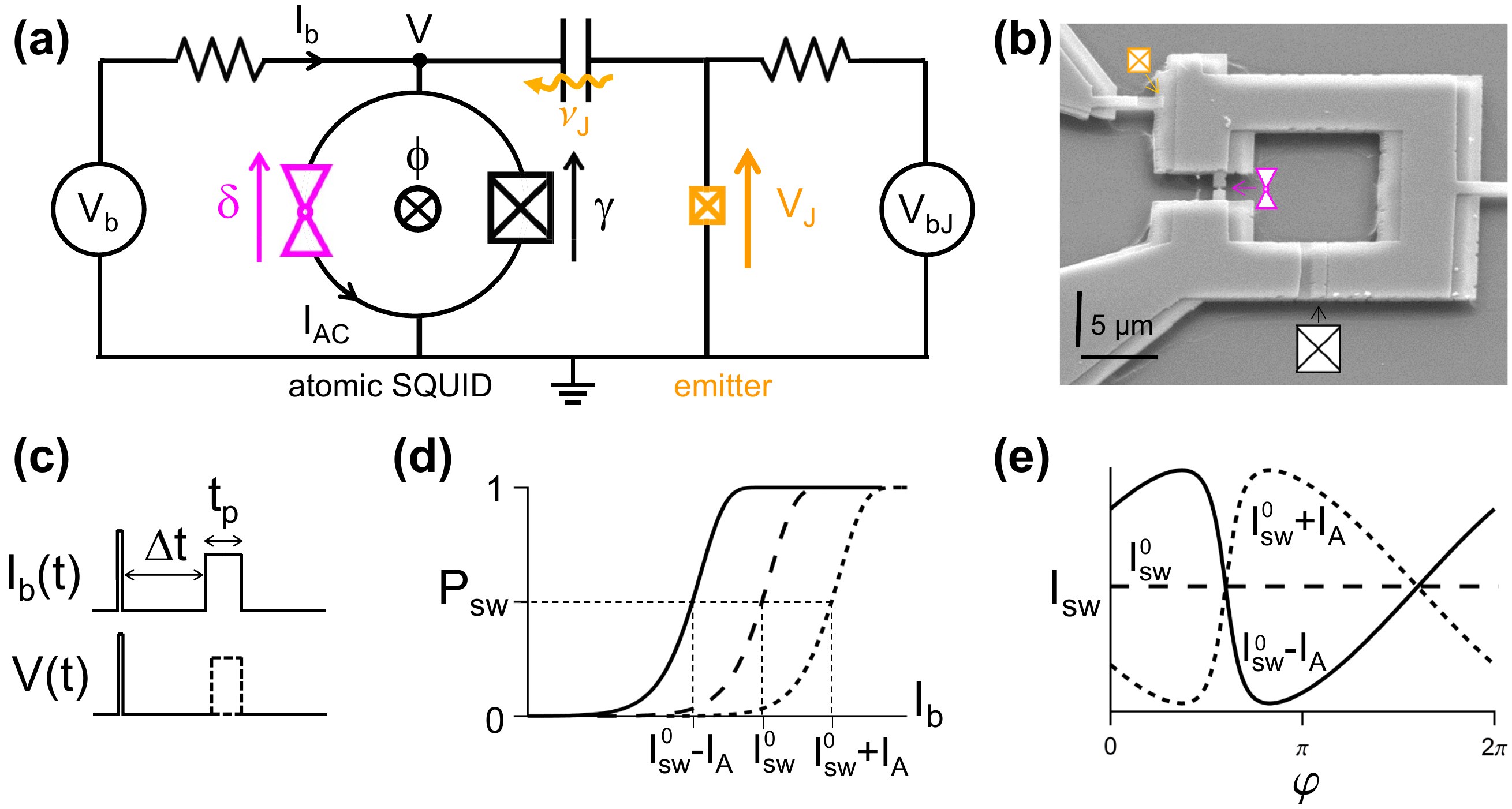}\caption{\textbf{(a}\textbf{\textcolor{black}{)}} Simplified schematic of the
experimental setup. An asymmetric SQUID is formed by an atomic contact
(magenta triangles) and an ancillary Josephson junction (critical
current $I_{0}\simeq1.06\,$\textmu{}A, 20 times larger than the typical
critical current of a one-atom aluminum contact). In the absence of
SQUID bias current $\left(I_{b}=0\right),$ magnetic flux $\phi$
threading the loop imposes a phase $\delta\simeq\varphi\equiv2\pi\phi/\phi_{0}$
across the contact and determines its excitation spectrum and the
loop current $I_{\rm{AC}}$. It is coupled through a capacitor to a voltage-biased
Josephson junction (yellow checked box, critical current $48\mathrm{\, nA}$)
used as an emitter: when biased at voltage $V_{J},$ it produces an
a.c. current at Josephson frequency $\nu_{J}=2eV_{J}/h$. \textbf{(}\textbf{\textcolor{black}{b)}}
Micrograph of the sample, seen under an angle (scale bars indicate
5~\textmu{}m in two directions). \textbf{(}\textbf{\textcolor{black}{c)}}
The SQUID switching current is measured by applying sequences of current
($I_{b}$) pulses of duration $t_{p}.$ A time $\Delta t$ before
each measurement pulse, any memory of the response to the previous
pulse is erased by a strong current pulse (``prepulse'') that forces
the SQUID to switch \cite{Zgirski}.\textbf{ }Switching events are
detected by the appearance of a voltage $V$ across the SQUID. \textbf{(}\textbf{\textcolor{black}{d)}}
Sketch of the switching probability $P_{\rm{sw}}$ of the atomic SQUID
as a function of the bias current $I_{b},$ for the atomic contact
in its ground state (solid line), in an odd state (dashed line) and
in the excited pair state (short-dashed line). For this figure, we
assumed a single channel and $I_{A}>0.$ The switching current of
the SQUID Josephson junction in the absence of an atomic contact is
$I_{\rm{sw}}^{0}.$\textbf{ (}\textbf{\textcolor{black}{e)}} Sketch of
the switching current of the SQUID as a function of the phase $\varphi$
in the ground, odd or excited even states.}

\label{setup} 
\end{figure*}

\section{EXPERIMENTAL SETUP}

The experimental setup is shown schematically in Fig.~\ref{setup}a.
An atomic contact is obtained by breaking in a controlled manner a
suspended constriction in an Al film $\left(\Delta=180\,\text{\textmu eV}\right)$
\cite{Ruitenbeek,LBThesis}. It is placed in parallel with a tunnel
Josephson junction having a much larger critical current $I_{0}\simeq1.06\,\mathrm{\text{\textmu A}}\gtrsim20\,\left|I_{A}\right|,$
hence forming an asymmetric SQUID. The number of conduction channels
in the atomic contact and their transmission probabilities are determined
from a fit of the current-voltage characteristic of the SQUID \cite{MLDR}.
The magnetic flux $\phi$ through the SQUID loop fixes the phase difference
at the atomic contact to $\delta=\varphi+\gamma$, with $\varphi=2\pi\phi/\phi_{0}$
the reduced flux, $\phi_{0}=h/2e$ the flux quantum, and $\gamma$
the phase across the SQUID junction. The SQUID is asymmetric enough
so that $\gamma\simeq\arcsin\left(I_{b}/I_{0}\right)$ is determined
only by the bias current $I_{b}.$ The SQUID is capacitively coupled
on-chip to a small Josephson junction (critical current $48\mathrm{\, nA}$),
called the ``emitter''. Figure~~\ref{setup}b shows a micrograph
of the SQUID and the emitter. The emitter is biased at a voltage $V_{J}$
and due to the a.c. Josephson effect acts as a broadband microwave
source at frequency $\nu_{J}=2eV_{J}/h.$ When $h\nu_{J}$ matches
a transition energy in the atomic contact, a photon can be absorbed,
and the occupation of Andreev states is modified. In ``absorption
spectroscopy'' experiments performed on the same device \cite{Nature},
energy dissipated during excitation of an Andreev transition in the
superconducting atomic contact was detected by measuring the d.c.
current through the emitter. Here instead, the corresponding change
in the supercurrent $I_{\rm{AC}}\left(\delta\right)$ of the atomic contact
is accessed by measuring the critical current of the SQUID: to a good
approximation, the critical current of such an asymmetric SQUID is
the sum of the critical current $I_{0}$ of the SQUID Josephson junction
and of $I_{\rm{AC}}\left(\delta\right).$

In practice, due to thermal fluctuations, the SQUID switches to a
finite voltage state at a current inferior to the critical current.
The current $I_{\rm{AC}}\left(\delta\right)$ is therefore inferred from
the switching probability $P_{\rm{sw}}\left(I_{b}\right)$ when a bias
current pulse of height $I{}_{b}$ and duration $t_{p}=1\,\text{\textmu s}$
is applied (Fig.~\ref{setup}c) \cite{quantronium}. The probability
$P_{\rm{sw}}\left(I_{b}\right)$ increases smoothly from 0 to 1 around
the switching current, which is $I_{\rm{sw}}^{0}$ for the junction alone
and $\sim I_{\rm{sw}}^{0}+I_{\rm{AC}}\left(\delta\right)$ for the atomic SQUID
\cite{MLDR}, \textit{i.e.} $I_{\rm{sw}}^{0}-I_{A}$ in the ground state,
$I_{\rm{sw}}^{0}+I_{A}$ in the excited state and $I_{\rm{sw}}^{0}$ in the
odd states (Fig.~\ref{setup}d\&e). The principle of our experiment
is to measure changes of $P_{\rm{sw}}$ induced by the microwave excitation.

\begin{figure*}[t]
\includegraphics[clip,width=1\textwidth]{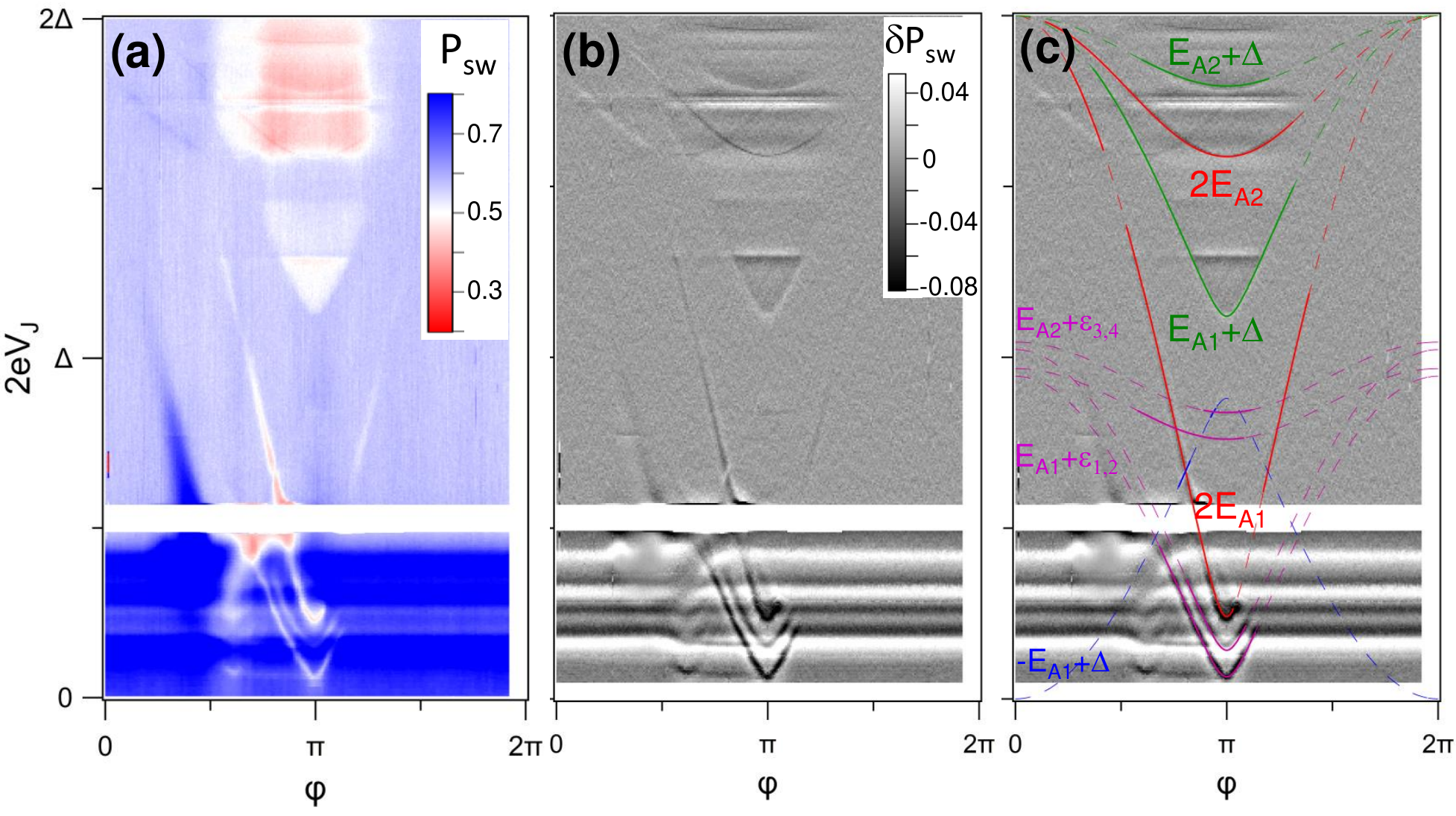}\caption{\textbf{(a) }Switching spectrum $P_{\rm{sw}}\left(\varphi,2eV_{J}\right)$
measured on an atomic contact with two channels of transmissions $0.985$
and $0.37.$ \textbf{(b)} Filtered spectrum $\delta P_{\rm{sw}}\left(\varphi,2eV_{J}\right)$
(see text) highlights rapid variations of the switching probability
with energy. \textbf{(c)} Same as (b), with transitions underlined.
Dashed lines correspond to the predicted positions of transitions,
and solid lines underline where they are actually visible in the data.
Colors correspond to those of arrows in Fig.~\ref{principle}a. In
red, Andreev transitions at $2eV_{J}=2E_{A1,2}$. Green lines: threshold
$2eV_{J}=E_{A1,2}+\Delta$ for transitions odd states with one quasiparticle
in an Andreev level, another one in the continuum. Blue line: threshold
$2eV_{J}=-E_{A1}+\Delta$ for transitions from odd states to even
states by transferring a quasiparticle from the Andreev level into
the continuum. Magenta lines: Andreev transitions induced by the second
harmonic of the excitation when $2eV_{J}=E_{A1,2}$, shifted by energies
$\varepsilon_{1-4}=-5,2,-3,4\,\text{\textmu}$eV (from bottom to top).}

\label{spectra} 
\end{figure*}

The switching probability is calculated from the response to a train
of $10^{4}$ pulses. For each value of the flux $\phi,$ the height
of the measurement pulse is set to $I_{\rm{sw}}$ such that $P_{\rm{sw}}=0.5$
in the absence of microwaves, \textit{i.e. }at $V_{J}=0$. Assuming
that the atomic contact is then in its ground state, $I_{\rm{sw}}=I_{\rm{sw}}^{0}-I_{A}\left(\delta\right)$
(Fig.~\ref{setup}d\&e). Then $P_{\rm{sw}}$ is recorded as a function
of the microwave excitation frequency set by $V_{J}$. This procedure
is repeated for all values of $\phi$, and one obtains a map of $P_{\rm{sw}}$
as a function of $\varphi$ and $V_{J}:$ the ``switching spectrum''.
Such a spectrum is shown in Fig.~\ref{spectra}a for a contact with
two channels of significant transmissions: $\tau_{1}=0.985$, $\tau_{2}=0.37$.
Most of the features of the rich spectrum, which has numerous
phase- and energy-dependent lines and plateaus, are explained below
and in the Supplemental Material \cite{supplemental}.

\section{INTERPRETATION OF THE SPECTRUM}

We first discuss general aspects of the spectrum which are unrelated
to Andreev physics. The white band slightly above $2eV_{J}\simeq0.5\Delta$
with no data corresponds to voltages that cannot be accessed due to
an instability in the I-V of the emitter when the Josephson frequency
matches the plasma frequency $\nu_{p}$ of the SQUID \cite{Nature}.
When $2eV_{J}>h\nu_{p},$ the background of the image is light blue,
corresponding to an overall increase of the switching probability
to $P_{\rm{sw}}\simeq0.55$ interpreted as an increased noise temperature
when the emitter is biased. When $2eV_{J}<h\nu_{p},$ the overall
increase in $P_{\rm{sw}}$ is even stronger $\left(P_{\rm{sw}}\simeq0.6-0.8\right),$
with broad, phase-independent stripes in the spectrum, corresponding
to a $10-20$~mK increase in the effective temperature, an effect
attributed to resonant activation during the measurement pulse \cite{resonant,sdependence}.

\subsection{Identification of the transition lines}

Relevant to the physics of Andreev levels are changes in $P_{\rm{sw}}$
that depend on the phase across the atomic contact. They are better
seen in Fig.~\ref{spectra}b, where the slow components of $P_{\rm{sw}}(V_{J})$
in Fig.~\ref{spectra}a have been filtered out to obtain $\delta P_{\rm{sw}}$
(the signal was first smoothed on $0.17\Delta$-intervals; the result
was then subtracted from the original spectrum). Three types of transitions
are resolved, corresponding to the arrows in Fig.~\ref{principle}a:
the Andreev transition, \textit{i.e.} the excitation of the Andreev
pair (red arrow); the transition to an odd state with a single quasiparticle
in the Andreev doublet, the second one being excited to the continuum
(green arrows); the excitation of a quasiparticle from the Andreev
level to the continuum (blue arrow).

Andreev transitions at $2eV_{J}=2E_{A}$ (red arrow in Fig.~\ref{principle}a)
are seen as sharp V-shaped lines centered at $\varphi=\pi,$ with
minima at $2E_{A1}\left(\pi\right)\simeq0.25\Delta$ for the channel
with transmission $\tau_{1}=0.985$ and $2E_{A2}\left(\pi\right)\simeq1.6\Delta$
for the channel with transmission $\tau_{1}=0.37$ (red lines in Fig.~\ref{spectra}c).
The variations across the spectra of the intensity of the lines are
discussed in the Supplemental Material \cite{supplemental}.

There are in addition two strong lines parallel to $E_{A1}\left(\delta\right)$
and two faint lines parallel to $E_{A2}\left(\delta\right)$ in the
spectrum (magenta lines in Fig.~\ref{spectra}c). They correspond
to exciting Andreev transitions with the second harmonic of the Josephson
frequency. We do not understand however why the lines are shifted
up or down with respect to the expected position $2eV_{J}=E_{Ai}$
by a few \textmu{}V.

Transitions from the ground state to an odd state with one quasiparticle
in the Andreev doublet (energy $E_{A}$), and another one at energy
larger than $\Delta$ in the continuum (green arrows in Fig.~\ref{principle}a
and green lines in Fig.~\ref{spectra}c) are best seen in the first
channel as a reduced $P_{\rm{sw}}$ (white in Fig.~\ref{spectra}a) in
a region defined by $2eV_{J}>E_{A1}+\Delta$. The corresponding threshold
for the second channel is also seen at $2eV_{J}=E_{A2}+\Delta$.

There is a faint transition at $2eV_{J}=-E_{A1}+\Delta$ (blue line
in Fig.~\ref{spectra}c). It corresponds to exciting a quasiparticle
from the Andreev doublet of the first channel to the continuum (blue
arrow in Fig.~\ref{principle}a). The detection of this odd-even
transition is explained if one assumes a finite probability that the
doublet is occupied in the absence of excitation (as already found
in former experiments \cite{Zgirski}). 

Blurred replica of the transition lines\textcolor{black}{{} are visible
shifted leftwards by $\sim0.4\pi.$ They correspond to transitions
induced not before, but during the measurement pulse, as the microwave
produced by the emitter is applied continuously. During the pulse,
a finite current flows through the SQUID junction and the phase across
the contact is no more $\delta=\varphi,$ but }$\delta=\varphi+\gamma_{\rm{sw}}$
\textcolor{black}{with $\gamma_{\rm{sw}}\simeq\arcsin\left(I_{\rm{sw}}^{0}/I_{0}\right)\sim0.4\,\pi.$
The replica of the odd-even transition discussed in the previous paragraph
is responsible for the sharp disappearance of the Andreev transition
line }at $\varphi\gtrsim1.04\pi,$ as discussed in the SM \cite{supplemental}.

\subsection{Sign of changes in $P_{\rm{sw}}$}

In most of the spectrum $\left(0.6\pi<\varphi\lesssim1.5\pi\right)$
the Andreev transitions and the transitions to odd states manifest
themselves by a decrease of the switching probability $P_{\rm{sw}}$ (white
on blue background in Fig.~\ref{spectra}a, black on grey background
in Fig.~\ref{spectra}b). In contrast, the faint odd-even transition
gives an increased $P_{\rm{sw}}.$ At $\varphi\sim0.6\pi,$ all transitions
disappear, and reappear at $\varphi<0.6\pi$ with opposite sign of
the change in $P_{\rm{sw}}$ (this effect is particularly visible on the
line at $2eV_{J}=2E_{A2}$ and on its replica).

The explanation is found in Fig.~\ref{setup}d\&e: starting from
the ground state (solid lines), all transitions lead to a decrease
of $P_{\rm{sw}}$ in the regions where $I_{A}\left(\delta\right)>0$. Since
$\delta=\varphi+\gamma_{\rm{sw}}$ during the measurement pulse, $I_{A}\left(\delta\right)>0$
when $0.6\pi<\varphi<1.6\pi.$

The odd-even transition (blue line) is seen because of a finite probability
to find initially the Andreev level in an odd state. Then the initial
pulse height, set such that $P_{\rm{sw}}=0.5$ when $V_{J}=0,$ corresponds
to a current between $I_{\rm{sw}}^{0}-I_{A}$ and $I_{\rm{sw}}^{0},$ and transitions
to the ground state cause an increase in $P_{\rm{sw}}.$

\subsection{Dynamics of the Andreev levels occupation during spectroscopy}

\begin{figure}
\includegraphics[clip,width=0.9\columnwidth]{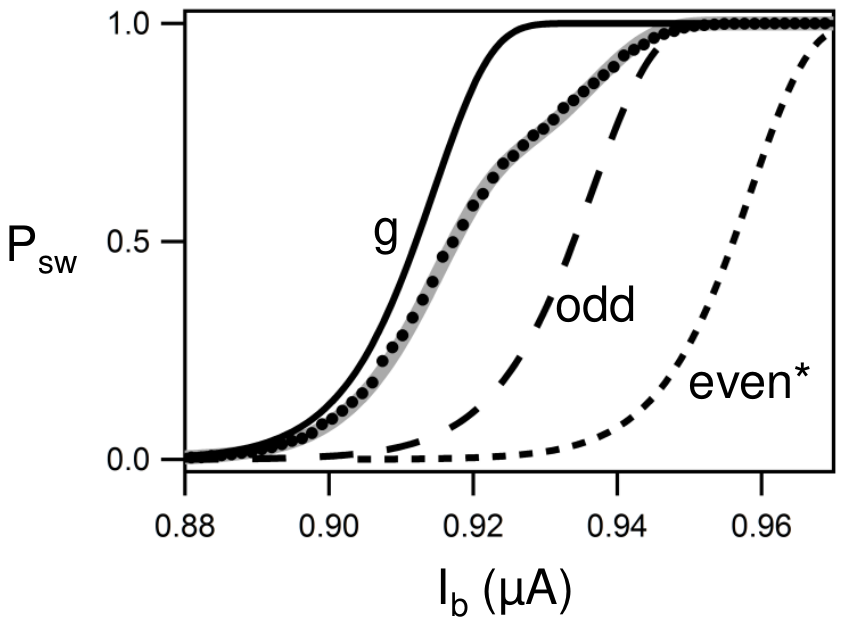}\caption{Dots: Switching probability $P_{\rm{sw}}\left(I_{b}\right)$ measured near
the minimum of the Andreev transition line at $2E_{A1}$ $(2eV_{J}=0.24\Delta,\,\varphi=1.02\pi)$.
Solid line: calculated $P_{\rm{sw}}\left(I_{b}\right)$ with the two channels
in the ground state (corresponding to a total current in the SQUID
junction $I_{b}-I_{A1}-I_{A2}$). Dashed line: calculated $P_{\rm{sw}}\left(I_{b}\right)$
with the first channel in an odd state (corresponding to a total current
in the SQUID junction $I_{b}-I_{A2}$). Grey curve: fit of the data
with sum of the two previous curves with weights 0.63 and 0.37. Short-dashed
line: calculated $P_{\rm{sw}}\left(I_{b}\right)$ with the first channel
in the even excited state (corresponding to a total current in the
SQUID junction $I_{b}+I_{A1}-I_{A2}$). A finite probability to find
the channel in the even excited state would have resulted in a contribution
of the short-dashed line to the data.}

\label{Scurve} 
\end{figure}

\begin{figure}
\includegraphics[clip,width=0.7\columnwidth]{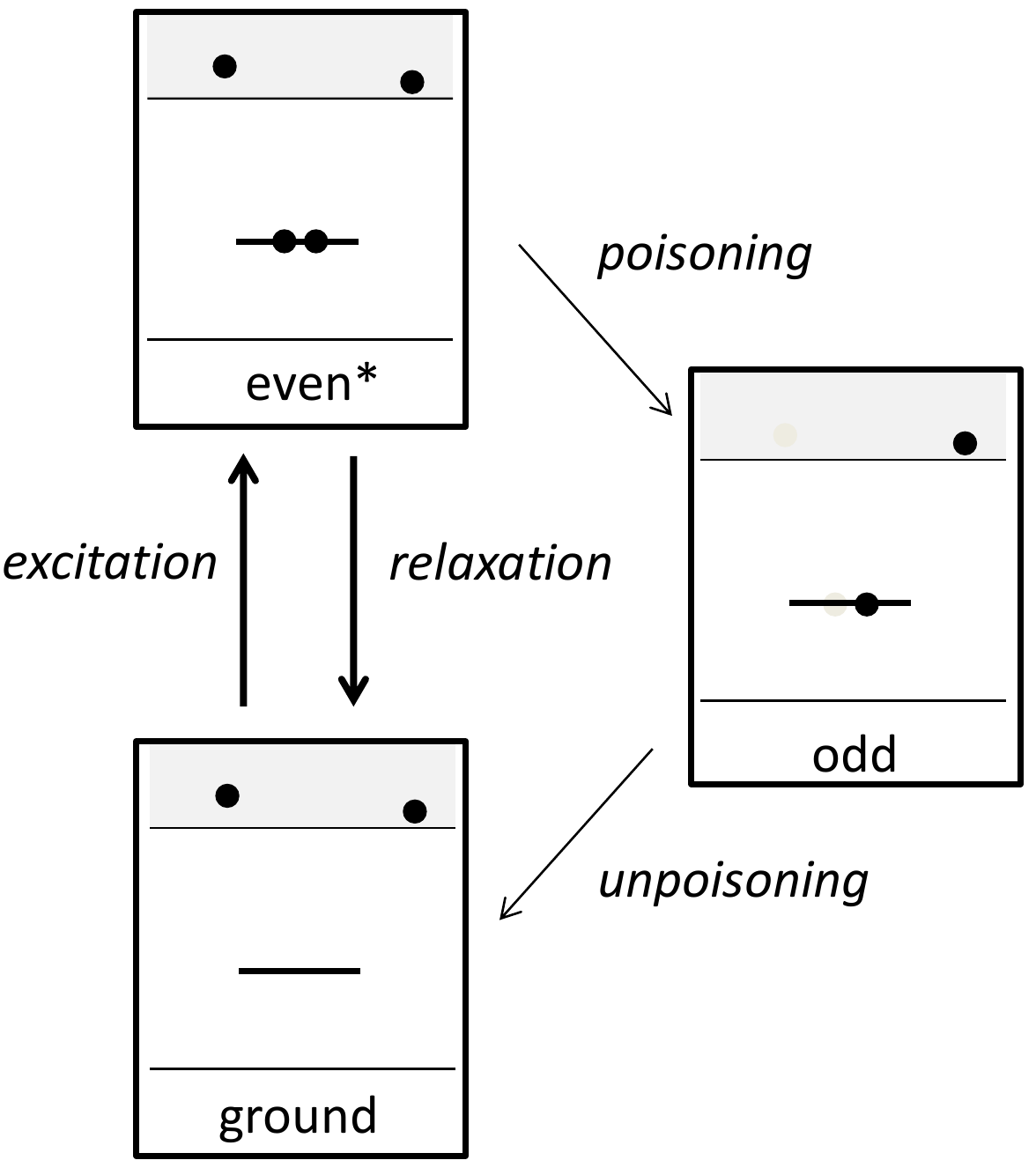}\caption{Dynamics of Andreev level occupation in presence of a resonant excitation
at frequency $2E_{A}.$ Andreev excitation are induced from the ground
state (bottom) to the excited even state (even{*}, top). When followed
by a direct relaxation, the process can immediately repeat itself,
and give rise to a d.c. current through the emitter junction, which
is the signal used for absorption spectroscopy \cite{Nature}. Alternatively,
one quasiparticle in the Andreev level can recombine with a quasiparticle
in the continuum, and lead to an odd state (poisoning). Such states
are long-lived, allowing the detection of the preceeding Andreev transition
by supercurrent spectroscopy. Odd states relax to the ground state
(unpoisoning) by a similar recombination process \cite{opposite}.}

\label{transitions} 
\end{figure}

We now analyze more in depth the amplitude of changes in $P_{\rm{sw}}$
and the nature of the states that are detected in the experiment.
When exciting Andreev transitions, one expects that $I_{\rm{AC}}\left(\delta\right)$
changes by $2I_{A},$ because the supercurrent carried by a channel
of transmission $\tau$ changes from $-I_{A}$ to $+I_{A}.$ This
turns out not to be the case: setting $\varphi$ and $V_{J}$ on the
Andreev transition line $2eV_{J}=2E_{A1}$, the curve $P{}_{\rm{sw}}\left(I_{b}\right)$
does not show the step corresponding to the excited state (see Fig.~\ref{Scurve}).
An explanation is sketched in Fig.~\ref{transitions}: when the Andreev
pair is excited, it often decays to its ground state; cycles of excitation/relaxation
give rise to a d.c. current through the emitter \cite{Nature}. But
it can also decay to an odd state (``quasiparticle poisoning''),
which is long-lived \cite{Zgirski}. Just before the measurement pulse,
the probability to be in an odd state can therefore be large.\textcolor{black}{{}
When the current increases through the SQUID, the phase across the
contact changes by up to $\gamma_{\rm{sw}},$ the Andr}eev energy changes
from $E_{A}\left(\varphi\right)$ to $E_{A}\left(\varphi+\gamma_{\rm{sw}}\right)$
and the excitation is no longer resonant with the Andreev transition.
The population of the excited state then decays at a rate $\Gamma_{r}.$
If $\Gamma_{r}>t_{r}^{-1}$ ($\Gamma_{r}>10\,$MHz) with $t_{r}\simeq0.1\,\text{\textmu}$s
the rise-time of the measurement pulse, it relaxes before the pulse
has fully developed and no change is detected in $P_{\rm{sw}}$ \cite{crossings}.
In contrast, the odd state is observed because its decay rate is in
general much smaller \cite{Zgirski,Diego}. Hence, quasiparticle poisoning
acts as a ``sample and hold'' mechanism for detecting Andreev transitions.
Note that transitions at energies greater than $E_{Ai}+\Delta$ (green
in Fig.~1a and 3c) leave the Andreev doublet directly in an odd state,
and their observation does not require quasiparticle poisoning.

For phases such that $\Delta-E_{A1}\left(\varphi+\gamma_{\rm{sw}}\right)<2E_{A1}\left(\varphi\right),$
a new process comes into play: unpoisoning by the microwaves during
the measurement pulse. Because the Andreev energy approaches $\Delta$
due to the phase shift $\gamma_{\rm{sw}},$ photons at energy $2E_{A1}\left(\varphi\right)$
can excite a quasiparticle from the Andreev level to the continuum
during the measurement pulse. This process empties the Andreev level
from an odd state and leads to the abrupt disappearance of the Andreev
transition line at $\varphi\gtrsim1.04\pi$ (see Fig.~8 in SM \cite{supplemental}).

When the Andreev doublet is in an odd configuration, energy absorption
at $2E_{A}$ is hindered, and the d.c. current through the emitter
associated to the absorption by the Andreev transition is suppressed.
This was probed in time-resolved measurements of the emitter current.
Quasiparticle poisoning and unpoisoning are then observed in real
time as a telegraphic noise in the current, with timescales in the
tens of ms (see Fig.~10 in SM \cite{supplemental}). 

\section{COMPARISON OF ABSORPTION SPECTROSCOPY AND SUPERCURRENT SPECTROSCOPY}

\begin{figure*}[t]
\includegraphics[clip,width=1\textwidth]{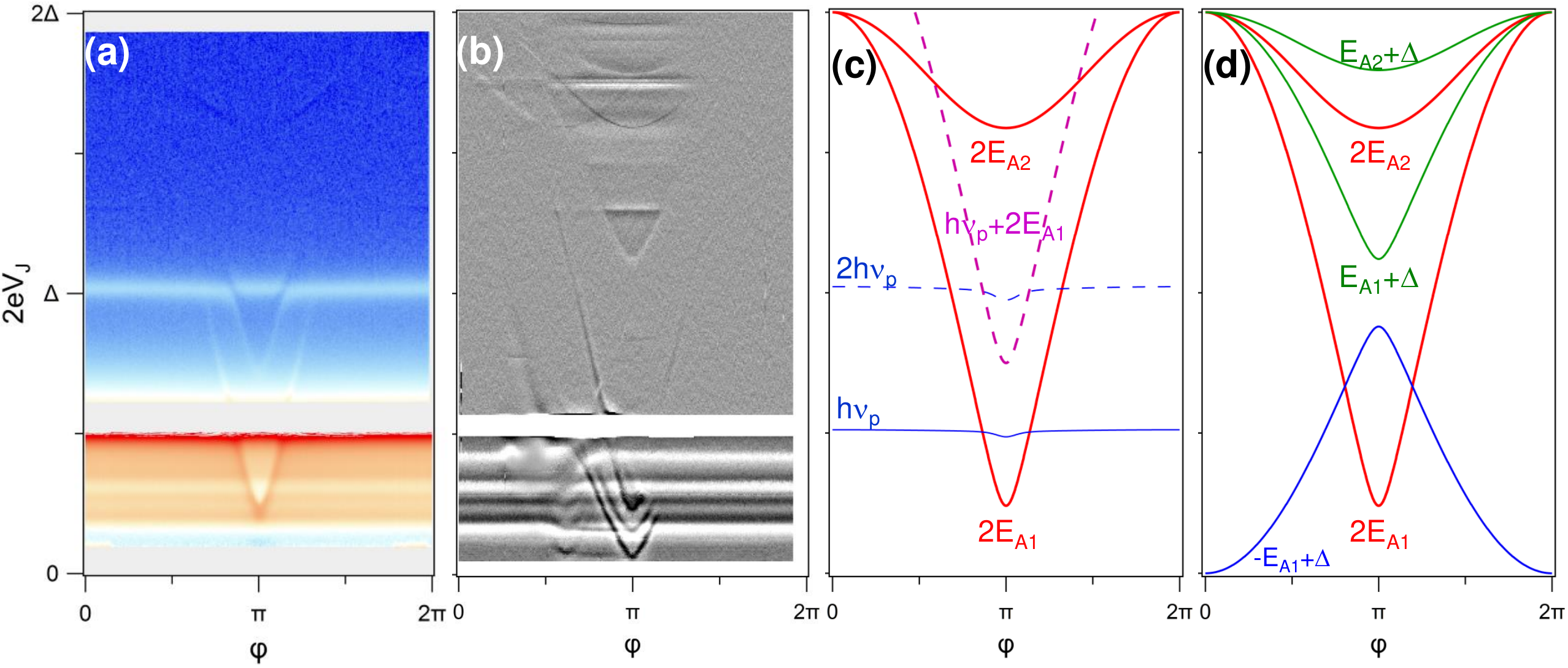}\caption{Comparison of the absorption spectrum \textbf{(a)} and of the switching
spectrum\textbf{ (b)} taken on the same contact ($\tau_{1}=0.985,\,\tau_{2}=0.37$).
\textbf{(c) }and \textbf{(d)} indicate the main transitions visible
in (a) and (b), respectively. Solid red in (c,d): Andreev transitions
at $2eV_{J}=2E_{A1,2}$. In (c): solid and dashed blue: plasma transition,
1st and 2nd harmonic ($2eV_{J}=h\nu_{p}$, $2h\nu_{p}$); dashed magenta
line: simultaneous excitation of Andreev and plasma modes ($2eV_{J}=2E_{A1}+h\nu_{p}$).
We also show in (d), with thin green lines, the threshold $2eV_{J}=E_{A1,2}+\Delta$
for simultaneous excitations of quasiparticles in the Andreev doublet
and in the continuum; thin blue line: threshold $2eV_{J}=-E_{A1}+\Delta$
for transitions from odd states to even states having one quasiparticle
in the continuum.}

\label{compspec} 
\end{figure*}

Finally, we compare in Fig.~\ref{compspec} the absorption spectrum
\cite{Nature} and the switching spectrum measured on the same atomic
contact. The spectra are different because absorption spectroscopy
detects excitation/relaxation cycles that repeat themselves rapidly
enough to give a substantial current through the emitter, whereas
supercurrent spectroscopy requires long-lived states that do not decay
during the measurement pulse rise time. Nevertheless, as explained
above, transitions to the short-lived even excited states are also
seen in supercurrent spectroscopy thanks to quasiparticle poisoning.

The switching spectrum contains more information, in particular in
the upper half of the spectrum. Remarkably, the Andreev transition
in the second channel $\left(2eV_{J}=2E_{A2}\right)$ which was barely
visible in the absorption spectrum is sharply resolved. The switching
spectrum does not include the lines associated with the plasma mode
of the SQUID at $2eV_{J}=h\nu_{p},$ $2h\nu_{p}$ and $h\nu_{p}+2E_{A1}$
(see Fig.~~\ref{compspec}c), which are apparent in the absorption spectrum. The
reason not to see the two first ones is that the lifetime of the plasma
mode, estimated to be in the ns range from the quality factor $Q\simeq22$
of the corresponding peak in the I-V of the emitter, is much shorter
than the rise time $t_{r}$ of the measurement pulse. However, one
would expect to detect the third transition, at $2eV_{J}=h\nu_{p}+2E_{A1}$,
which corresponds to a simultaneous excitation of the plasma mode
and of the Andreev doublet. We speculate that even if poisoning occurs
in the same manner as when $2eV_{J}=2E_{A1},$ photons at energy $h\nu_{p}+2E_{A1}$
trigger unpoisoning (see SM \cite{supplemental}) and the doublet
is found only in its ground state \cite{eval}.

\section{CONCLUSIONS}

The experimental results show that spectroscopy based on the measurement
of the Josephson supercurrent allows detecting single quasiparticle
excitations in superconducting weak links. The entire excitation spectrum
is explained by the presence of Andreev doublets that can be occupied
by 0, 1 or 2 quasiparticles. Andreev transitions are detected when
followed by quasiparticle poisoning, which acts as a ``sample and
hold'' mechanism by placing the Andreev doublet in a long-lived odd
state. We also demonstrate for the first time the possibility to induce,
without injecting any charge \cite{Pillet}, transitions from the
(even) ground state to an odd state with a single excitation in the
Andreev doublet (the second one being in the continuum). This type
of transition could be used to prepare spin-qubits based on odd states
\cite{Chtchelkatchev,Padurariu}.

We thank Alfredo Levy Yeyati, John Martinis and Eva Dupont-Ferrier
for discussions. We gratefully acknowledge help from other members
of the Quantronics group, in particular P. Senat and P. F. Orfila
for invaluable technical assistance. Work partially financed by ANR
through projects DOC-FLUC, MASH, and by CNANO-Ile-de-France. The research
leading to these results has received funding from the People Programme
(Marie Curie Actions) of the European Union's Seventh Framework Programme
(FP7/2007-2013) under REA grant agreement n\textdegree{} PIIF-GA-2011-298415. 

\section{SUPPLEMENTAL MATERIAL}

\label{specsupp-1} 

\subsection*{Current-phase relation}

In Figure 7, we show the measured switching current
of the atomic SQUID, which reveals the current-phase relation of the
atomic contact in its ground state \cite{MLDR}.

\begin{figure}[h!]
\includegraphics[width=0.8\columnwidth]{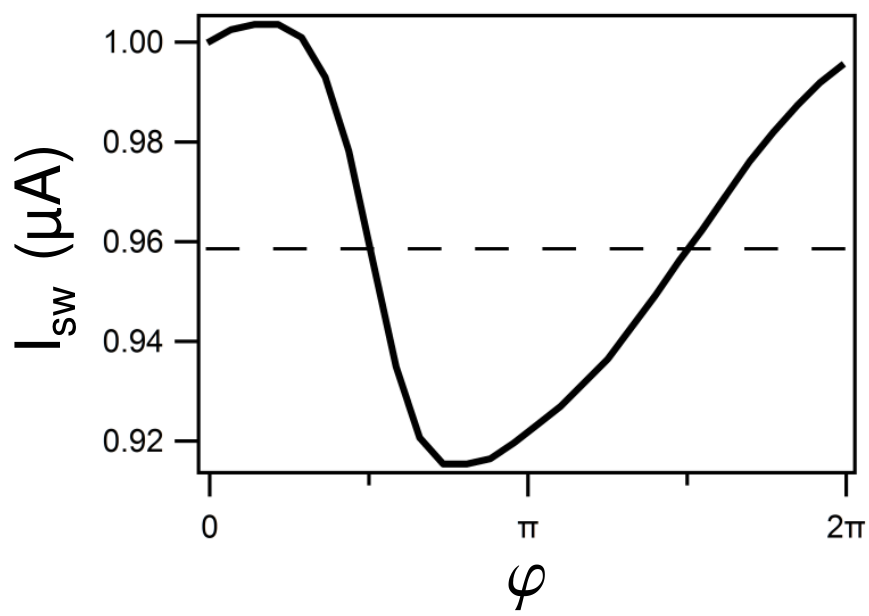}\caption{Switching current $I_{{\rm {sw}}}\simeq I_{{\rm {sw}}}^{0}+I_{{\rm {AC}}}\left(\varphi+\gamma_{{\rm {sw}}}\right)$
of the SQUID in absence of excitation, defined by $P_{{\rm {sw}}}\left(I_{{\rm {sw}}}\right)=0.5$
when $V_{J}=0,$ as a function of reduced flux $\varphi.$ The horizontal
line marks $I_{{\rm {sw}}}^{0}.$ }
\end{figure}

\subsection*{Visibility of the transition lines}

\begin{figure*}[t!]
\includegraphics[width=1\textwidth]{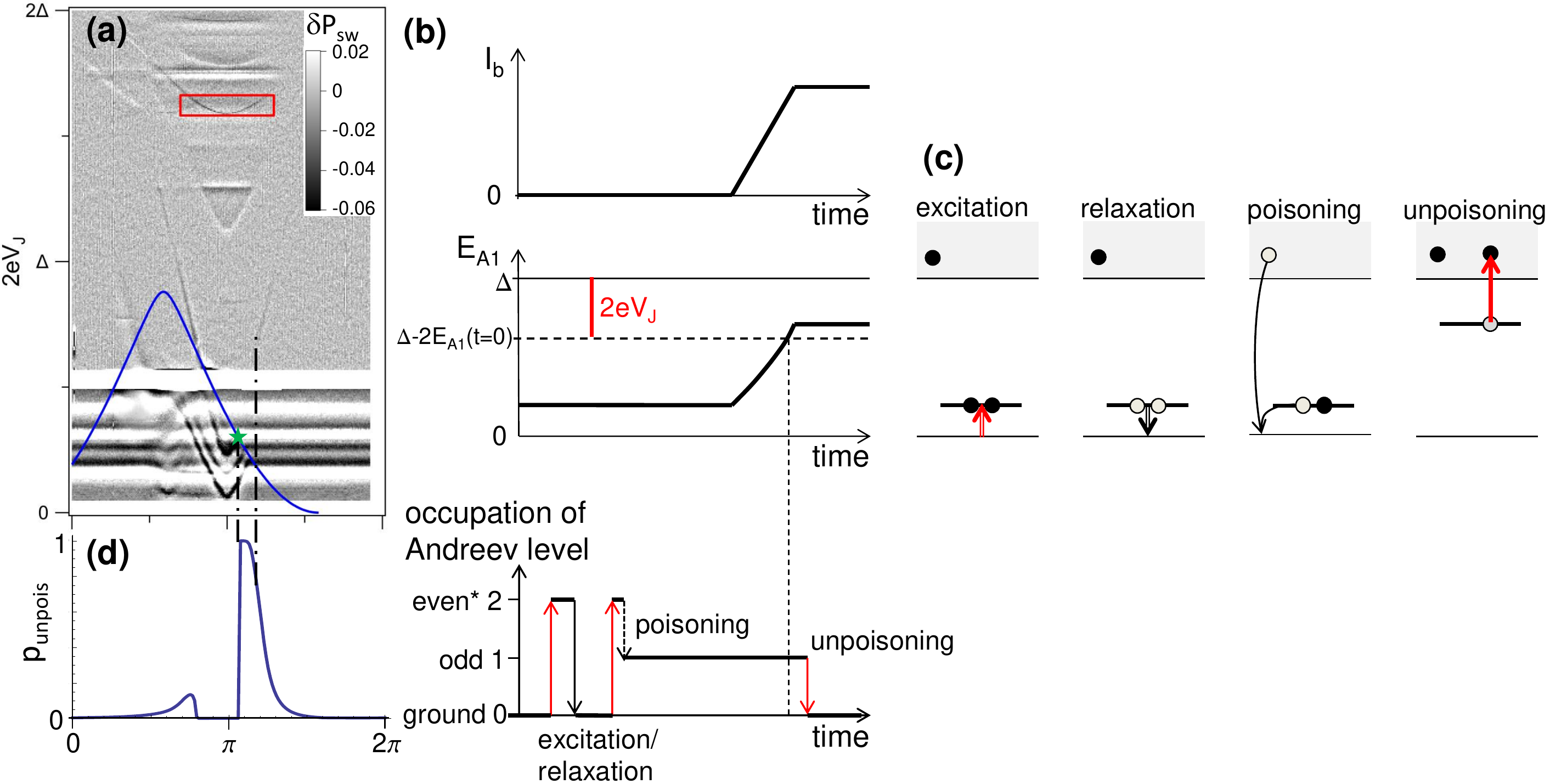}\caption{\textbf{(a)} Switching spectrum and threshold $2eV_{J}=\Delta-E_{A1}(\varphi+\gamma_{{\rm {sw}}})$
(blue line) for the excitation of a quasiparticle from the Andreev
level to the continuum during the measurement pulse (red box indicates
region zoomed in Fig.~9). The green star signals the position where
the Andreev transition line $2eV_{J}=2E_{A1}$ suddenly disappears.
\textbf{(b)} Sketch of the evolution of the system when biased resonantly
with the Andreev transition, at a position just next to the green
star, above the blue line $(2eV_{J}=2E_{A1}\left(1.1\pi\right)).$
Top: bias current of the SQUID, just before and at the beginning of
a measurement pulse; middle: time evolution of the Andreev energy
$E_{A1},$ which changes because the phase across the contact is $\delta=\varphi+\gamma_{{\rm {sw}}}\simeq\varphi+\arcsin(I_{b}/I_{0});$
bottom: typical evolution of the occupation of the Andreev level.
\textbf{(c)} Processes at play. Before the measurement pulse, the
system cycles between ground and even excited state, through excitation
(induced by the excitation at energy $2E_{A1}$) and relaxation processes.
At some point, poisoning occurs: the Andreev doublet becomes singly
occupied. During the measurement pulse, $E_{A1}$ increases, and when
it reaches $\Delta-2E_{A1}(t=0)$ (vertical dashed line) the energy
of the photons from the emitter is sufficient to excite the quasiparticle
from the Andreev level to the continuum, and the Andreev system can
return to its ground state (unpoisoning).\textbf{ (d)} Numerical evaluation
of $p_{{\rm {unpois}}},$ the unpoisoning probability during measurement. }

\label{specsupp} 
\end{figure*}

Andreev transitions manifest themselves by a decrease of the SQUID
switching probability $P_{{\rm {sw}}}$ for $\varphi\gtrsim0.5\pi,$
and by an increase of $P_{{\rm {sw}}}$ for $\varphi\lesssim0.5\pi$.
They disappear on the right-hand side of the spectrum, for $\varphi\gtrsim1.4\pi.$
The change in the sign of the effect is related both to the shape
of the current-phase relation of the atomic contact {[}Fig.~7 and
Fig.~2(e){]}, which leads to variations of the SQUID switching current
around $I_{{\rm {sw}}}^{0}$, and to the shift of the phase across
the atomic contact by $$\gamma_{{\rm {sw}}}\simeq\arcsin\left(I_{b}/I_{0}\right)\sim0.4\pi$$
when the current pulse is applied (the equality is exact only in the
limit $I_{A}/I_{0}\rightarrow0$). At $\varphi=\pi,$ the symmetry
point for the system in the absence of current in the SQUID, $I_{A}=0$
so that the current through the atomic contact does not depend on
the occupation of the ABS. The transition line is nevertheless visible
because when the measurement current pulse is applied, the phase $\delta$
shifts to $\varphi+\gamma_{{\rm {sw}}}\simeq1.4\pi,$ a phase at which
$I_{A}$ is positive. This is seen in the ground state in Fig.~7,
where $I_{{\rm {sw}}}\left(\varphi=\pi\right)-I_{{\rm {sw}}}^{0}=-I_{A}\left(1.4\pi\right)\simeq-0.04\,$\textmu{}A.
Any transition is therefore expected to increase the switching current,
hence decreasing the measured $P_{{\rm {sw}}}\left(I_{b}\right)$
for fixed measurement pulse height, in agreement with the data. Similarly,
transitions induced at $\varphi\sim0.5\pi$ while $I_{b}=0$ are invisible
because $\varphi+\gamma_{{\rm {sw}}}$ is close to $\pi$ during the
measurement pulse, and the current through the atomic contact is zero
for all configurations, as seen in Fig.~7 where $I_{{\rm {sw}}}\left(\varphi=0.5\pi\right)\sim I_{{\rm {sw}}}^{0}.$
This explains why the sign of the changes in $P_{{\rm {sw}}}$ is
that of $I_{{\rm {sw}}}\left(\varphi\right)-I_{{\rm {sw}}}^{0}.$
Finally, the absence of signal for $1.5\pi\lesssim\varphi<2\pi$ corresponds
to the fact that when the measurement pulse is swept up, $\varphi$
approaches or crosses $2\pi,$ so that Andreev levels merge with the
continuum where the quasiparticles decay away.

Another remarkable feature of the data is the abrupt disappearance
of the $2eV_{J}=2E_{A1}$ Andreev transition line when $\varphi\gtrsim1.04\pi$
(green star in Fig.~\ref{specsupp}(a)). In this region, microwaves
of energy $2eV_{J}$ can excite quasiparticles from the Andreev doublet
to the continuum during the measurement pulse {[}see Fig.~\ref{specsupp}(b)
and Fig.~\ref{specsupp}(c){]}. Then, the doublet is restored to
its ground state and $P_{{\rm {sw}}}$ is unaffected. Quantitatively,
this requires that the rate $W_{5}$ for this process (using the notations
of Ref.~\cite{Kos}) is larger than the switching rate of the SQUID
when the first channel is in the odd state. To evaluate $W_{5}$,
two ingredients are needed: the real part of the corresponding admittance
$\mathrm{Re}Y_{5}^{(0)}\left(\omega\right)$ given in Ref.~\cite{Kos}
and the amplitude of the phase excitation $\delta_{ac}$ due to the
emitter. The phase excitation depends on the ratio $x=2eV_{J}/h\nu_{p}(s)$
of the excitation energy and the plasma frequency $\nu_{p}(s)=\nu_{p}^{0}\left(1-s^{2}\right)^{1/4}$
of the current-biased SQUID ($s=I_{{\rm {sw}}}^{0}/I_{0}\simeq0.95$):
$$\delta_{ac}=\left(LI_{J}^{0}/\varphi_{0}\right)/\left|1-x^{2}+jx/Q\right|.$$
Here, $I_{J}^{0}\simeq48\,$nA is the critical current of the emitter
junction, $L\simeq0.18\,$nH is the parallel combination of the SQUID
junction inductance $\varphi_{0}/I_{0}$ and of the inductance of
the biasing circuit \cite{LBThesis} and $Q\approx22$ is the quality
factor of the plasma frequency, evaluated from the line shape of the
corresponding peak in the I-V of the emitter junction. At $x=0,$
$\delta_{ac}\simeq0.03\,$rad. Using $$W_{5}=\delta_{ac}^{2}\frac{\hbar}{2e^{2}}\omega\mathrm{Re}Y_{5}^{(0)}\left(\omega\right)$$
\cite{Kos}, one obtains the unpoisonning probability $p_{{\rm {unpois}}}$
of the odd state during the measurement pulse (including the rise
time), see Fig.~\ref{specsupp}(d). This explains why the Andreev
lines disappear abruptly when $\varphi\gtrsim1.04\pi,$ corresponding
to the threshold for exciting a quasiparticle from the Andreev level
to the continuum, and reappears at $\varphi\gtrsim1.2\pi$ because
the rate $W_{5}$ diminishes below the MHz range. Note that the same
unpoisoning process, which is in principle always possible when $E_{A}>\Delta/3$
\cite{Kos}, has a negligible probability when $2eV_{J}=2E_{A2}.$

\subsection*{Dynamics}

\begin{figure}[tbh]
\includegraphics[clip,width=1\columnwidth]{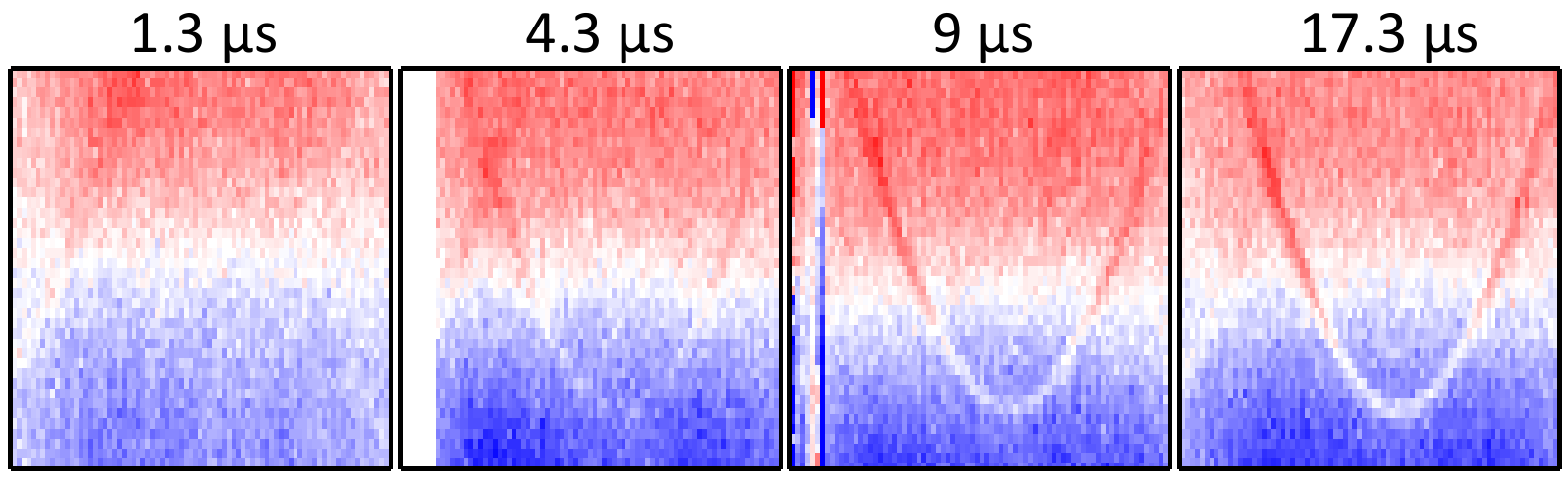}\caption{Spectroscopy Andreev transition of the second channel $\left(2eV_{J}=2E_{A2}\right)$
near $\varphi=\pi$, with increasing delay $\Delta t$ between prepulse
and measurement pulse (1.3 to 17.3~\textmu{}s). In each sub-spectrum,
the x-axis spans the interval $0.7\pi<\varphi<1.3\pi,$ the y-axis
$1.58\Delta<2eV_{J}<1.66\Delta$ (red box in Fig.~8(a)).}

\label{timedep} 
\end{figure}

The dynamics of the occupation of the Andreev doublet sketched in
Fig.~5 was explored for the second channel by varying the delay $\Delta t$
between prepulse and measurement pulse (Fig.~\ref{timedep}). The
poisoning rate is too weak to observe transitions to the odd state
for short delays ($\Delta t=1.3\,$\textmu{}s) but becomes apparent
in a characteristic time of $9\,$\textmu{}s. 

Absorption spectroscopy measurements on another contact probed the
real-time fluctuations between the odd and even occupations of a channel
with transmission 0.97 at $\varphi=\pi.$ The current-voltage characteristics
of the emitter junction shows bistability around $V_{J}=32\,\text{\textmu}$V,
the voltage corresponding to the Andreev transition at $\varphi=\pi$
(top panel in Fig.~10). The corresponding telegraphic noise on the
current (measured with a 5~kHz bandwidth), corresponds to rates of
15~Hz (down rate) and 53~Hz (up rate) (bottom panel of Fig.~10).

\begin{figure}[tbh]
\includegraphics[clip,width=0.8\columnwidth]{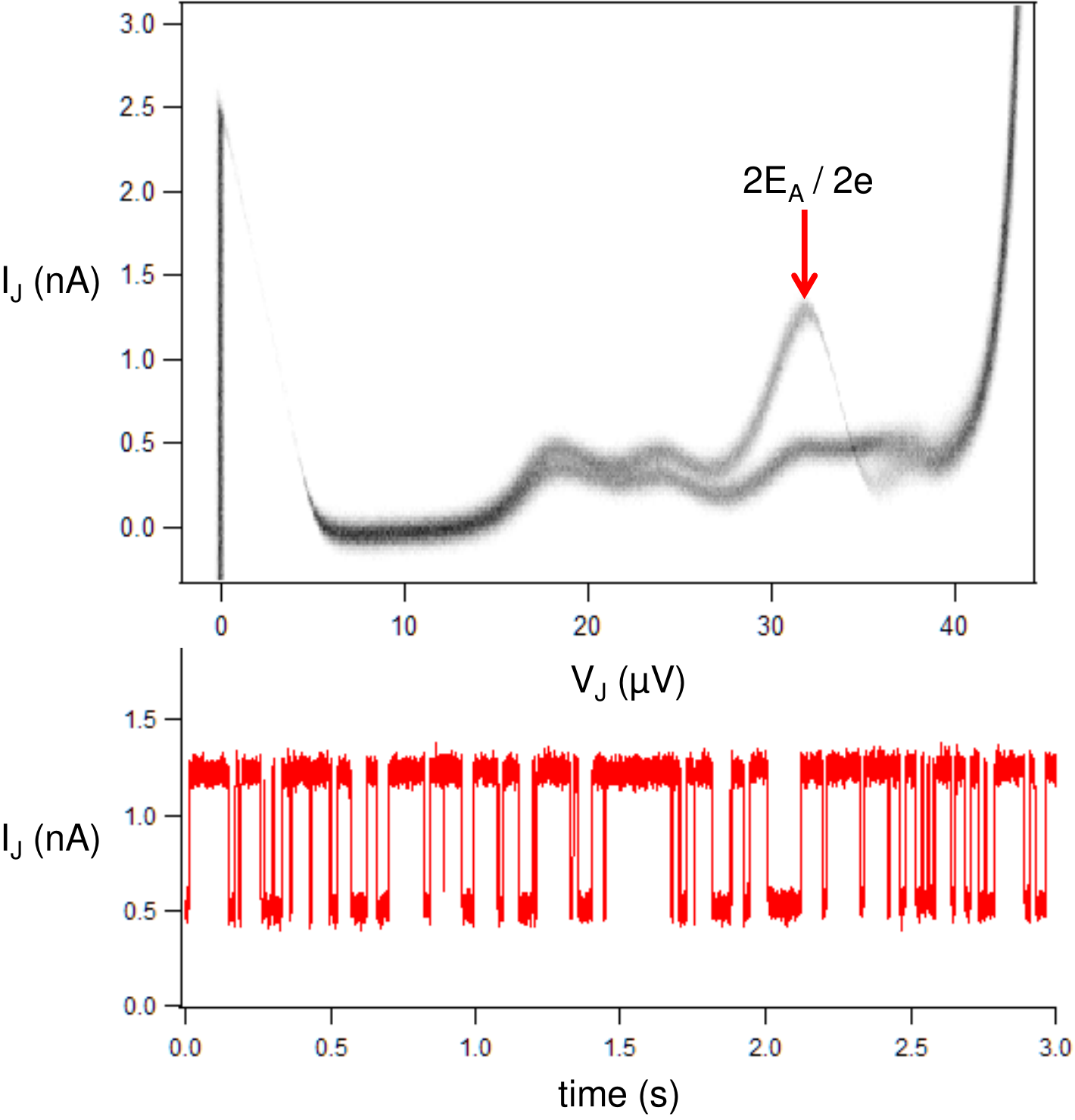}\caption{Top: Multiple traces of the current-voltage characteristic of the
emitter junction coupled to an atomic contact with a channel with
transmission 0.97, at $\varphi\sim\pi,$ showing bistability. Bottom:
time dependence of the current $I_{J}$ through the emitter biased
at $V_{J}=2E_{A}/2e$ (red arrow in top panel). The current is high
when the system cycles between ground and even excited state, and
low in an odd state.}

\label{timedep-1} 
\end{figure}


\begin{thebibliography}{10}
\bibitem{SNS}Such junctions, with have a large number of Andreev
levels, are described quasi-classically in terms of a continuum of
sub-gap states and a continuous occupation function.
\bibitem{sdependence}Measurements on another contact showed that
this structure shifted to lower voltages when longer measurement pulses
were applied. This is compatible with the current dependence of the
plasma frequency $\nu_{p}\propto\left(1-I_{b}/I_{0}\right)^{1/4}$,
since switching occurs at smaller values of $I_{b}$ when longer pulses
are used.
\bibitem{crossings}Moreover, when $0.55\pi<\varphi<1.15\pi,$ $2E_{A1}\left(\varphi+\gamma\right)$
crosses the plasma energy $h\nu_{p}\left(I_{\rm{sw}}\right)$ during the
rise-time of the pulse. Since the excited Andreev state and the plasma
mode are strongly coupled \cite{Nature}, and since the lifetime of
the plasma mode is in the ns range, this crossing causes relaxation
to the ground state.
\bibitem{opposite}Transitions in the opposite direction are not represented
because their rates are much smaller \cite{Zgirski,Diego}.
\bibitem{eval}Although numerical evaluations of the rate for this
unpoisoning process using Ref.~\cite{Kos} are an order of magnitude
too small to validate this hypothesis.
\end{thebibliography}
\end{document}